\documentclass[a4paper,11pt]{article}
\usepackage{pos}
\usepackage{bm}
\newcommand{\pa}{\partial}
\newcommand{\nn}{\nonumber}
\title{Transport coefficients of heavy mesons in a thermal medium}

\author*[a]{Juan M. Torres-Rincon}

\affiliation[a]{Departament de F\'isica Qu\`antica i Astrof\'isica and Institut de Ci\`encies del Cosmos (ICCUB), Facultat de F\'isica,  Universitat de Barcelona, Mart\'i i Franqu\`es 1, 08028 Barcelona, Spain}

\emailAdd{torres@fqa.ub.edu}

\abstract{

We have investigated the many-body equations of $D$ and $\bar{B}$ mesons in a thermal medium by applying an effective field theory based on chiral and heavy-quark symmetries. Exploiting these symmetries within kinetic theory, we have derived an off-shell Fokker-Planck equation which incorporates information of the full spectral functions of these states.

In this contribution we present our latest results on heavy-flavor transport coefficients below the chiral restoration temperature, in both charm and bottom sectors. The calculation incorporating temperature-dependent spectral functions and interactions, together with off-shell effects, allows for an improved matching to the state-of-the-art calculations above the chiral transition temperature. }

\FullConference{%
  41st International Conference on High Energy physics - ICHEP2022\\
  6-13 July, 2022\\
  Bologna, Italy
}

\begin{document}
\maketitle

\section{Introduction: effective field theory at finite temperature}

Within the so-called hard probes, heavy-flavor observables play an important role to gain knowledge of the initial stages of relativistic heavy-ion collisions~\cite{Song:2015sfa}. Heavy quarks evolve in a rapidly changing medium interacting with light partons in a similar manner to what a relativistic Brownian particle does in a locally thermalized medium. These heavy particles eventually hadronize into $D$ and $\bar{B}$ mesons (among other heavy states), and interact with light hadrons, leaving the medium at the kinetic freeze-out~\cite{He:2022ywp}.

In previous works~\cite{Tolos:2013kva,Montana:2020lfi,Montana:2020vjg} we have developed an effective field theory to describe these heavy mesons interacting with light mesons at low energies. Interactions, upon the unitarization of the perturbative scattering amplitudes, provide valuable information about the bound and scattering states which are generated by the nonperturbative $T$-matrix~\cite{Guo:2009ct,Liu:2012zya}. In Ref.~\cite{Montana:2020lfi,Montana:2020vjg} this effective theory was extended to finite temperature by applying the imaginary-time formalism. The method allows to quantify the medium modification of the ground states $D, D^*, {\bar B}, {\bar B}^*$. Notice that the formalism considers the $D^*, {\bar B}^*$ as stable heavy-quark spin partners of the $D, {\bar B}$ mesons, respectively. 

\begin{figure}[ht]
\centering
\includegraphics[width=1\textwidth]{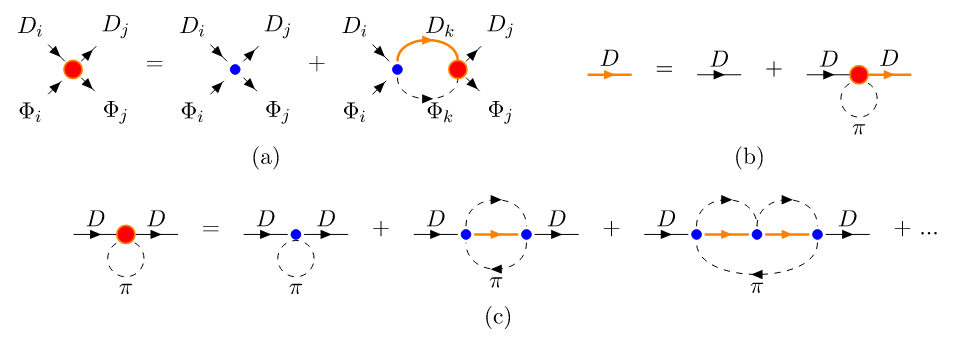}
\caption{Self-consistent set of diagrams to be computed at finite temperature. (a) $T$-matrix equation with a medium-modified heavy-meson propagator, (b) $D$-meson propagator dressed by the $T$-matrix interaction with particles of the bath (e.g. pions), (c) $D$-meson self-energy upon expansion of the $T$-matrix.}
\label{fig:eqfinT}
\end{figure}

At $T\neq 0$ within the imaginary-time formalism the tree-level perturbative amplitude is the same as in vacuum. It is derived analytically at next-to-leading order and can be found in Refs.~\cite{Montana:2020lfi,Montana:2020vjg}. The key step to obtain a sensible description of the interactions, is to enforce the unitarity of the scattering-matrix amplitudes. Applying the on-shell factorization into the Bethe-Salpeter equation one can express the unitarized $s-$wave amplitude $T$, in terms of the perturbative amplitude $V$~\cite{Tolos:2013kva,Montana:2020lfi,Montana:2020vjg},
\begin{equation}
T_{ij}(E,\bm{P}) = V_{ij}(E,\bm{P}) + \sum_k V_{ik} (E,\bm{P}) \ G_k(E,\bm{P}) \ T_{kj}(E,\bm{P}) \ , 
\end{equation}
where $i$ and $j$ are the incoming and outgoing scattering channels, $E$ and $\bm{P}$ are the total energy and momentum in the center-of-mass frame, and the two-meson loop function reads,
\begin{align}
\label{eq:hot-loop-compact}
 G(E,\bm{P}\,;T)  =\int\frac{d^3q}{(2\pi)^3}\int_{-\infty}^\infty d\omega\int_{-\infty}^\infty d\omega'&\frac{S_1(\omega,\bm{q}\,;T)S_2(\omega',\bm{P}-\bm{q}\,;T)}{E-\omega-\omega'+i\varepsilon} \left[1+f(\omega;T)+f(\omega';T)\right] \ .
\end{align}
In Eq.~(\ref{eq:hot-loop-compact}) thermal modifications enter in the spectral functions, $S_i(\omega,\bm{q};T)=-\textrm{Im} [\omega^2-\bm{q}^2-m_i^2-\Pi_i(\omega, \bm{q}; T]^{-1}/\pi$, and in the Bose-Einstein distribution functions, $f(\omega;T)$. In our scheme the light meson spectral function is approximated by the vacuum one, while the $D$-meson one (and $\bar{B}$ meson) needs to be self-consistently computed via the self-energy and the $T$-matrix,
\begin{equation} 
\Pi_D(\omega, \bm{q};T) = \frac{1}{\pi} \int \frac{d^3 q'}{(2\pi)^3} \int dE \frac{\omega}{\omega_l} \frac{f(E;T) - f(\omega_l;T)}{\omega^2-(\omega_l-E)^2+ i \epsilon \textrm{sgn} (\omega)} \ \textrm{Im } T_{Dl} (E, \bm{q}+\bm{q'}; T) \ ,
\end{equation}
where $\omega_l^2=q'^2+m_l^2$, with $l$ denoting the light meson (e.g. pion). The self-consistent scheme relating the $T$-matrix, $D$-meson propagator, and its self-energy is schematically shown in Fig.~\ref{fig:eqfinT}.

\begin{figure}[ht]
\centering
\includegraphics[width=0.42\textwidth]{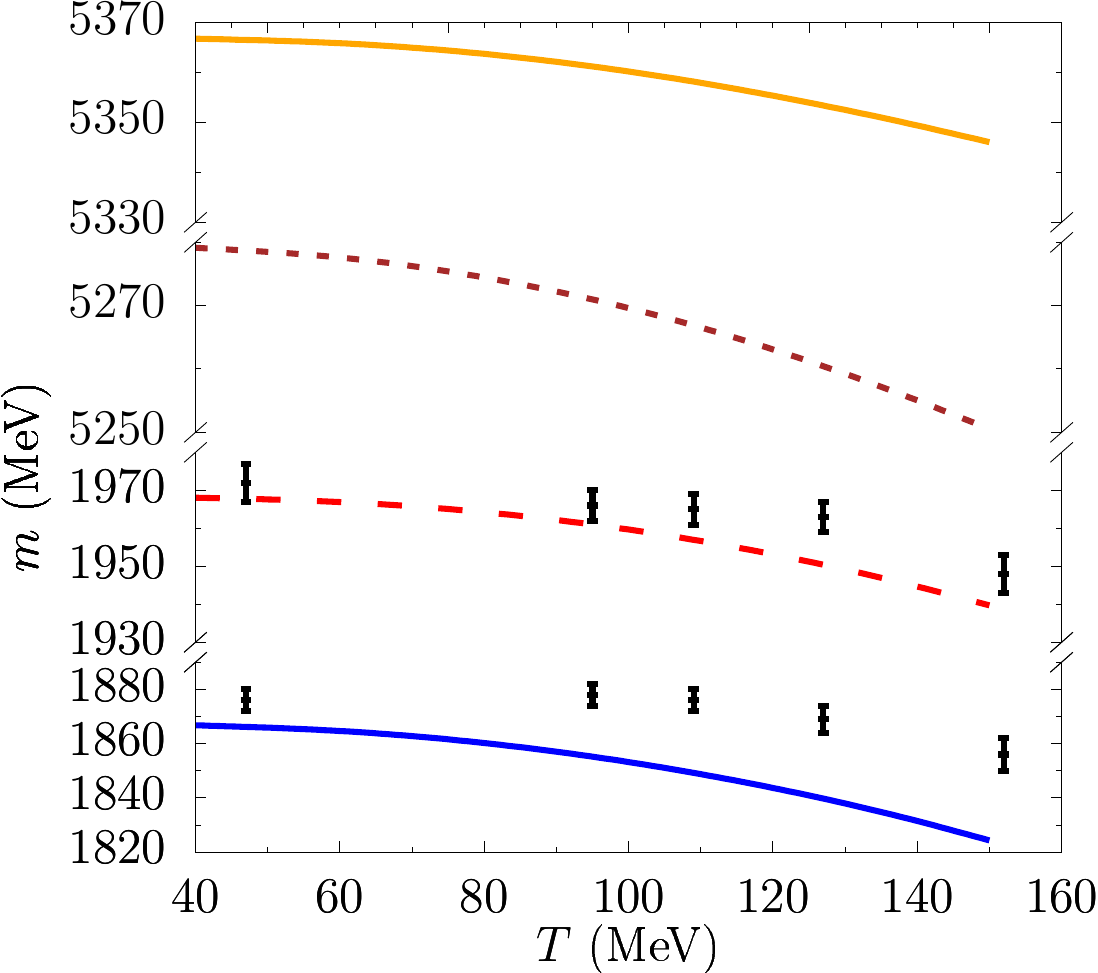}
\hspace{5mm}
\includegraphics[width=0.4\textwidth]{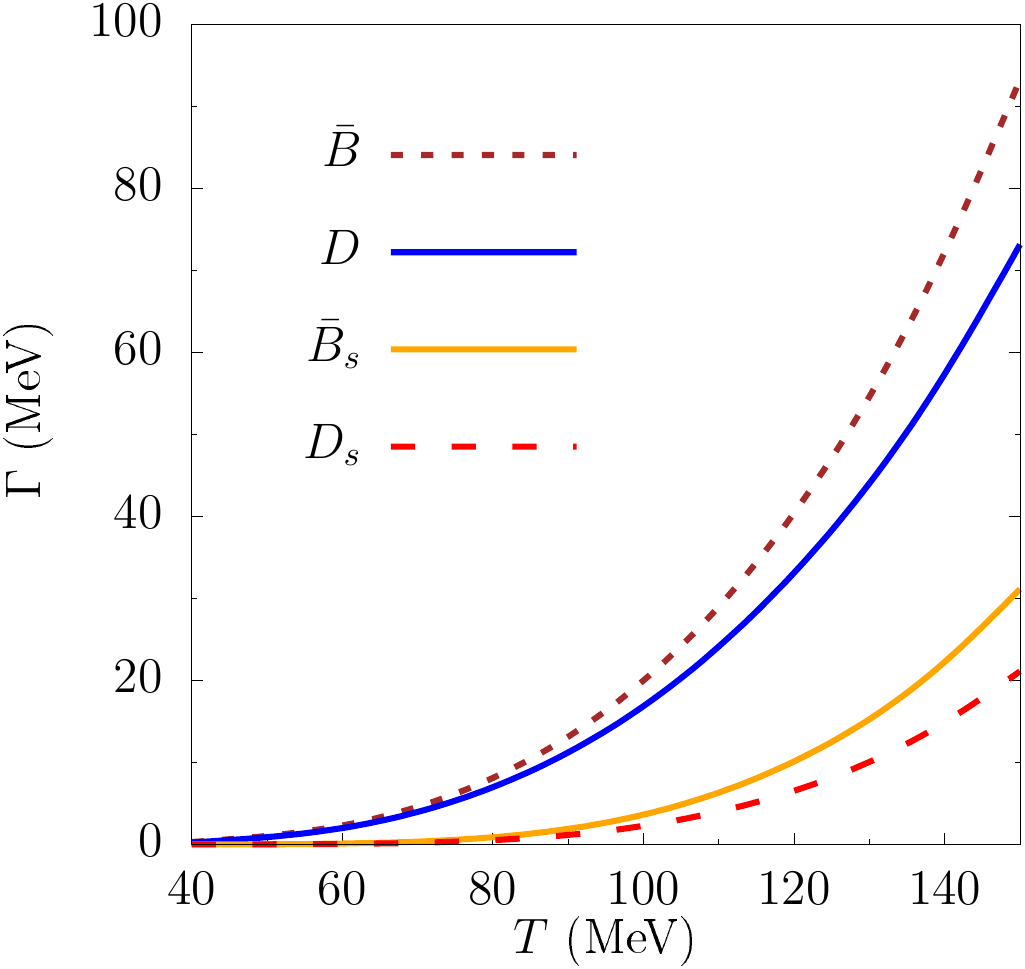}
\caption{Thermal masses (left) and decay widths (right) of the ground states of charm and bottom mesons. Lattice-QCD results taken from Ref.~\cite{Aarts:2022krz} are also shown for the masses of $D$ and $D_s$ mesons in the left panel.}
\label{fig:masses}
\end{figure}

Once the system is solved one can read off the spectral function of the heavy meson. We obtain that for all temperatures the decay width is very small compared to the thermal mass. Therefore, the $D$-meson always behaves as a narrow quasiparticle. In particular, the thermal masses and decay widths of the $D, D_s, \bar{B}, \bar{B}_s$ mesons are shown in the left and right panels of Fig.~\ref{fig:masses} (vector mesons can be computed similarly). Thermal masses of $D$ and $D_s$ mesons computed in the lattice-QCD calculation of~\cite{Aarts:2022krz} are also shown (a systematic shift in the lattice-QCD results should be expected due to the heavier-than-physical pions used in~\cite{Aarts:2022krz}). 

\section{Kinetic theory: off-shell Fokker-Planck equation}

More recently we have computed the transport coefficients of heavy mesons incorporating their quasiparticle nature, i.e. their off-shell properties acquired by the presence of the thermal medium. While the transport coefficients themselves are defined in equilibrium, it is necessary to detour and construct the non equilibrium kinetic equation, in order to obtain their precise expressions~\cite{ChapmanCowling}.

In Ref.~\cite{Torres-Rincon:2021yga} we developed the Kadanoff-Baym equation for heavy mesons in the so-called $T$-matrix approximation~\cite{kadanoff1962quantum}, which is the natural extension of the diagrammatic resummation performed at equilibrium when the scattering amplitudes were unitarized. The resulting transport equation with off-shell effects is then expanded for collisions where the exchanged momentum is small. The final kinetic equation is an off-shell Fokker-Planck equation~\cite{Torres-Rincon:2021yga},

\begin{equation}
\frac{\pa}{\pa t} G_D^< (t,k) = \frac{\pa}{\pa k^i} \left\{ \hat{A} (k;T) k^i G_D^< (t,k) + \frac{\pa}{\pa k^j} \left[ \hat{B}_0(k;T) \Delta^{ij} + \hat{B}_1(k;T) \frac{k^i k^j}{{\bm k}^2} \right] G_D^< (t,k) \right\} \ , \label{eq:offFP} 
\end{equation}
where $G_D^< (t,k^0,\bm{k})$ is the Wigner function of the heavy meson in a spatially homogeneous and isotropic medium. It is a function of time, and of energy $k^0$ and momentum $\bm{k}$ separately. The projector $\Delta^{ij}=\delta^{ij}-k^ik^j/\bm{k}^2$ and the (off-shell) transport coefficients read,
\begin{align}
 \hat{A} (k^0, {\bm k};T) & \equiv \left \langle 1 -\frac{{\bm k} \cdot {\bm k}_1}{{\bm k}^2} \right \rangle \ , \label{eq:hatA} \\
 \hat{B}_0 (k^0, {\bm k};T) & \equiv  \frac14 \left \langle {\bm k}_1^2 - \frac{({\bm k} \cdot {\bm k}_1)^2}{{\bm k}^2} \right \rangle \ , \label{eq:hatB0} \\
  \hat{B}_1 (k^0, {\bm k};T) & \equiv \frac12 \left \langle  \frac{[ {\bm k} \cdot ({\bm k}-{\bm k}_1)]^2}{{\bm k}^2} \right \rangle \ , \label{eq:hatB1}
\end{align}
where the brackets define the following thermal average
\begin{align}
\left\langle {\cal F}({\bm k},{\bm k}_1) \right\rangle & = \frac{1}{2k^0} \sum_{\lambda,\lambda'=\pm} \lambda \lambda' \int_{-\infty}^\infty \ dk_1^0 \int \prod_{i=1}^3 \frac{d^3k_i}{(2\pi)^3} \frac{1}{2E_22E_3}  \ S_D(k_1^0,{\bm k}_1)   \nn \\
& \times (2\pi)^4 \delta^{(3)} ({\bm k}+{\bm k}_3-{\bm k}_1-{\bm k}_2) \delta (k^0+\lambda' E_3- \lambda E_2-k^0_1) |T(k^0+ \lambda' E_3,{\bm k}+{\bm k}_3)|^2  \nn \\
& \times f(\lambda'E_3;T) \ [1+f (\lambda E_2;T)  ] \ [1+f (k_1^0;T)] \ \ {\cal F}({\bm k},{\bm k}_1)  \ . \label{eq:rateoff}
\end{align}

\begin{figure}[ht]
\centering
\includegraphics[width=0.42\textwidth]{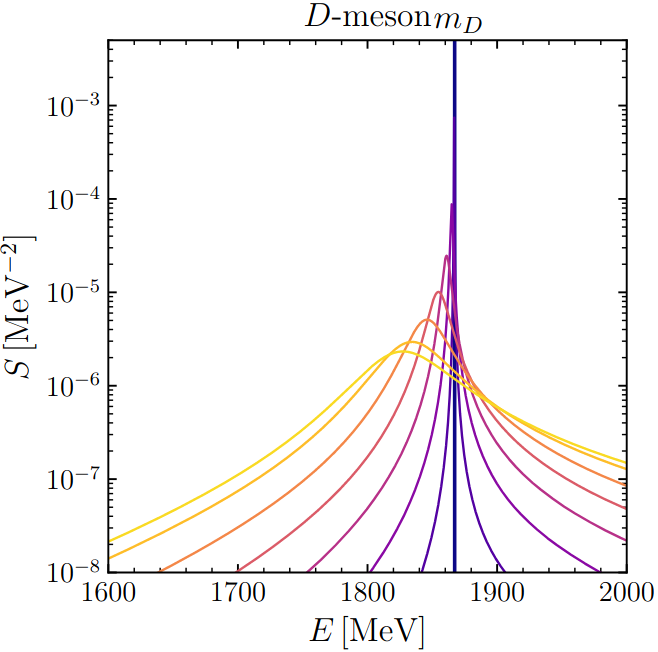}
\includegraphics[width=0.11\textwidth]{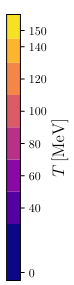}
\hspace{5mm}
\includegraphics[width=0.42\textwidth]{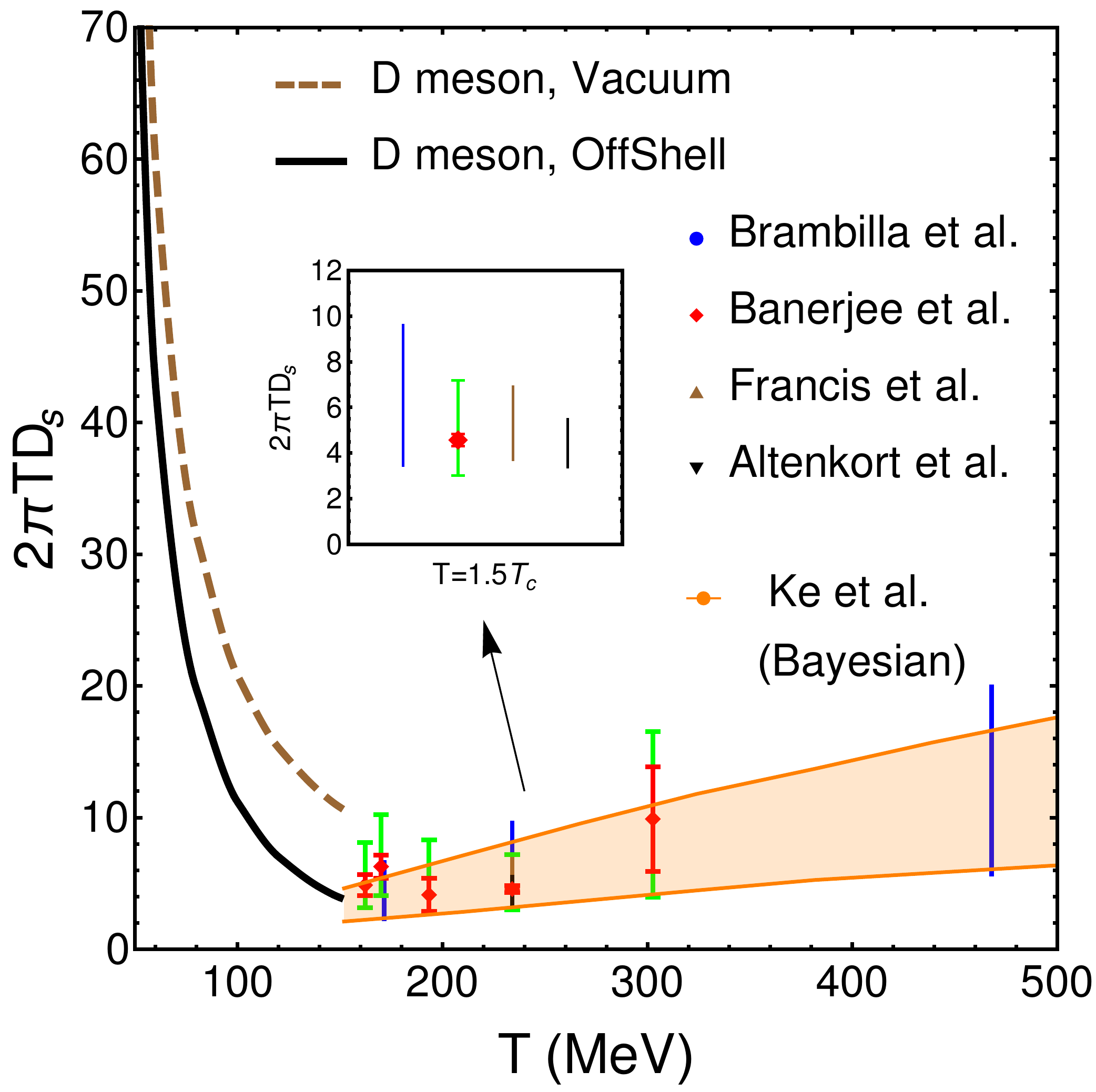}
\caption{Left: Spectral function of a $D$ meson at rest as a function of its energy and temperature. Right: Charm diffusion coefficient as a function of temperature computed for $D$ mesons in the low-temperature phase, and at high temperatures by several other works. See Ref.~\cite{Torres-Rincon:2021yga} for details and references.}
\label{fig:spectralDs}
\end{figure}

The on-shell limit of the previous equations reduces to the well-known expressions used in the past~\cite{Tolos:2013kva}. In the present case there exist several additions which improve the final results: 1) The equilibrium spectral function is introduced at finite temperature, and it is not a Dirac delta anymore, but a (narrow) distribution (see left panel of Fig.~\ref{fig:spectralDs}); 2) The scattering $T$-matrices also contain finite-temperature corrections; 3) New kinematical processes appear at finite temperature which are forbidden in vacuum~\cite{Weldon:1983jn}. These can be classified by analyzing the values of $\{ \lambda, \lambda' \}=\pm 1$ in Eq.~(\ref{eq:rateoff}). When $\lambda = - \lambda'$ the scattering is a $1\leftrightarrow 3$ process in which an off-shell heavy-meson decays/generates due to the interaction with the medium. The case $\lambda=\lambda'$ describes two binary collision processes. One is the standard heavy-light meson scattering evaluated above the unitary threshold of the $T$-matrix (and the only contribution which survives at $T=0$). The other one is a $2\leftrightarrow 2$ process dominated by the $T$-matrix evaluated below threshold (specifically at the so-called Landau cut of the 2-meson loop function $G(E,\bm{P})$~\cite{Weldon:1983jn}). This last scattering process is only present at $T \neq 0$, and for moderate temperatures cannot be neglected. The kinematics and physical meaning of each scattering process will be detailed in a future publication.

\section{Results and conclusions: charm and bottom diffusion coefficients}

In Ref.~\cite{Torres-Rincon:2021yga} we present the transport coefficients $\hat{A},\hat{B}_0$ and $\hat{B}_1$ in the soft momentum (static) limit, evaluated at the quasiparticle peak $k^0=E_k \equiv \sqrt{k^2+m_D^2(T)}$. In that reference we have shown how at low temperatures the off-shell transport coefficients coincide with those computed without any medium modification (since off-shell effects get largely suppressed). However, at moderate temperatures these effects start contributing sizably, and at our highest temperature $T= 150$ MeV the new kinematic processes for the heavy-meson relaxation make the transport coefficients increase a factor of 2-3 with respect to their values using vacuum interactions. 

In this contribution we only show the spatial diffusion coefficient $D_s(T)$. Within the classical (non relativistic) picture it can be interpreted as the average diffusive speed of heavy particles, and can be computed as $D_s(T)=T^2/B_0(k^0=E_k, \bm{k} \rightarrow 0; T)$. For the charm case ($D$ mesons) it is shown in the right panel of Fig.~\ref{fig:spectralDs} together with several lattice-QCD calculations, and a Bayesian analysis of real heavy-ion collisions at high energies (see Ref.~\cite{Torres-Rincon:2021yga} for details). Our result at low temperatures denoted as ``OffShell'' contains all thermal and off-shell effects. It supersedes the previous result (marked as ``Vacuum'' in the figure) computed with vacuum interactions. It can be seen how our latest calculation makes a better job in matching the high-$T$ predictions around the crossover temperature $T_c \simeq 156$ MeV.

\begin{figure}[ht]
\centering
\includegraphics[width=0.43\textwidth]{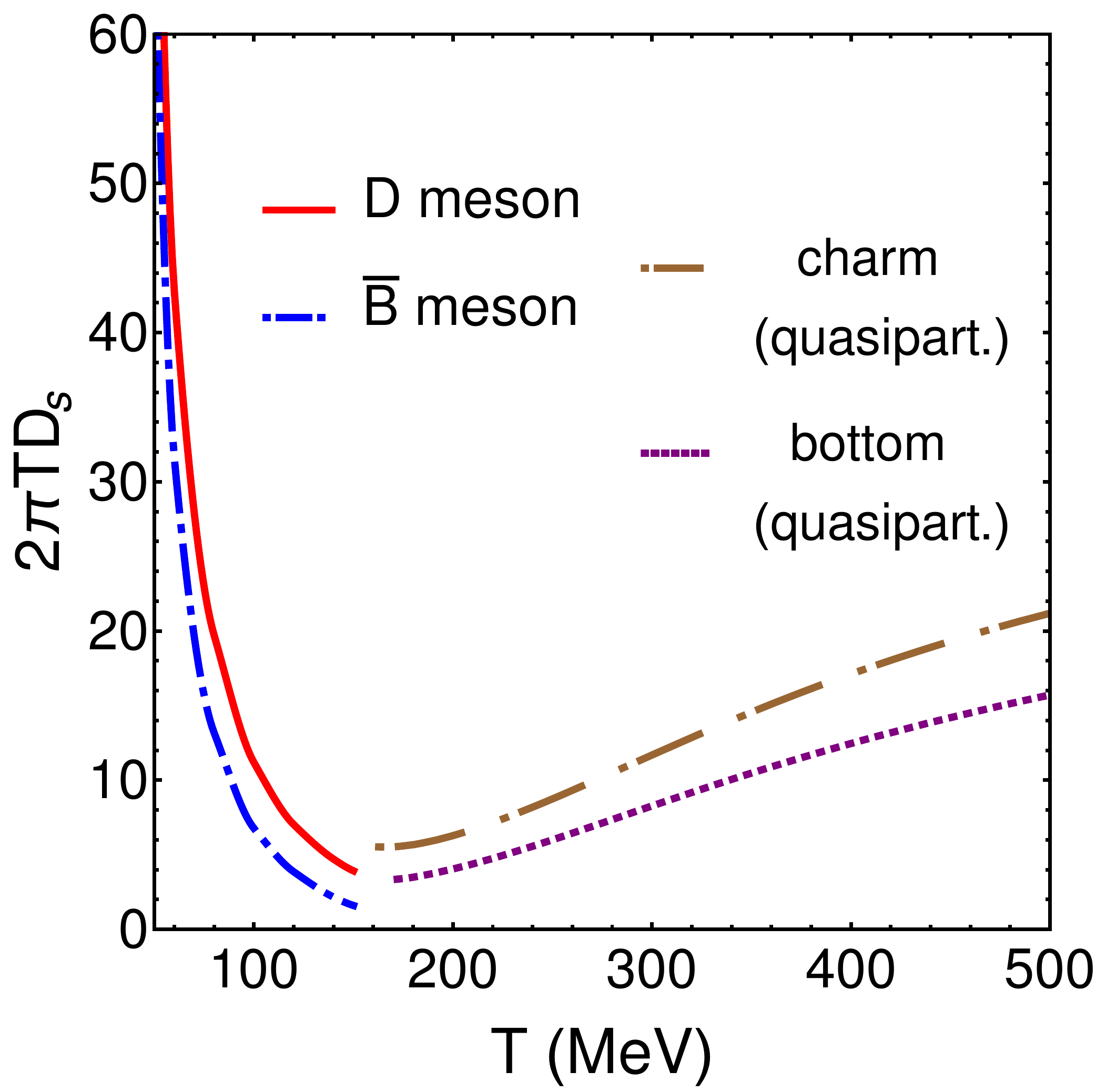}
\hspace{5mm}
\includegraphics[width=0.43\textwidth]{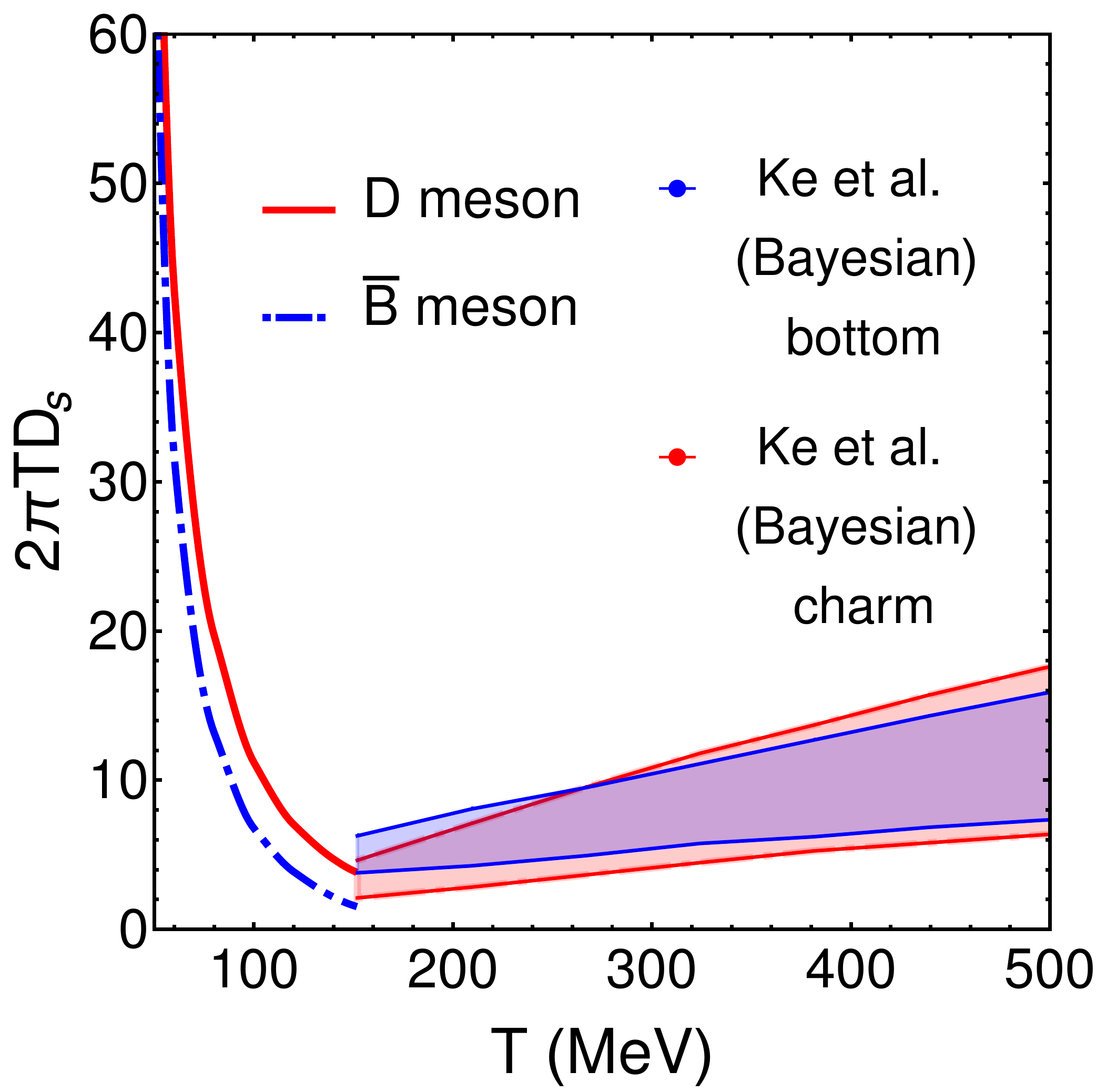}
\caption{Spatial diffusion coefficient of charm and bottom flavors as computed in the low-temperature phase by us, and in the high-temperature limit by the quasiparticle model of~\cite{Das:2016llg} for $c$ and $b$ quarks (left panel), and using Bayesian analyses of high-energy heavy-ion collisions~\cite{Ke:2018tsh} (right panel).}
\label{fig:DsComp}
\end{figure}

We finally compare our results for charm and bottom by computing the spatial diffusion coefficient of $D$ and $\bar{B}$ mesons. We obtain a similar temperature behavior with a slightly smaller diffusion coefficient for $\bar{B}$ mesons, see Fig.~\ref{fig:DsComp}. This result is consistent with the quasiparticle model of~\cite{Das:2016llg} in the high-temperature side (left panel of Fig.~\ref{fig:DsComp}), but not with the Bayesian results~\cite{Ke:2018tsh} which discriminate charm and bottom flavor coefficients (right panel of Fig.~\ref{fig:DsComp})---although the big error band does not allow to get a solid conclusion on the flavor-ordering of the diffusion coefficient. Notice that the non relativistic kinetic theory predicts that the dependence of $D_s$ on the heavy mass at leading order in density goes like $
D_s \propto \sqrt{1+m_l/m_H} \  $~\cite{ChapmanCowling},
which decreases with increasing $m_H$ (but it is finite for $m_H \rightarrow \infty$) like it happens in our case.

\acknowledgments
The results reported here were obtained in collaboration with Gl\`oria Monta\~na, \`Angels Ramos and Laura Tol\'os, to whom I would like to express my gratitude for the fruitful collaboration. This project was financed by the Spanish MCIN/ AEI/10.13039/501100011033/, the EU STRONG-2020 project under the program  H2020-INFRAIA-2018-1 grant agreement no. 824093, and the German DFG through projects no. 411563442 (Hot Heavy Mesons) and no. 315477589 - TRR 211 (Strong-interaction matter under extreme conditions).

\end{document}